\documentclass[journal]{IEEEtran}
\include{top_misc}
\usepackage{cite}

\usepackage{epstopdf}
\usepackage{graphicx}
\usepackage{url}
\usepackage{amsmath,amssymb,amsfonts}
\usepackage{subfigure}
\usepackage{wrapfig}
\usepackage{fancyhdr}
\renewcommand{\thispagestyle}[1]{} 

\newtheorem{theorem}{Theorem}[section]

\newtheorem{Definition}[theorem]{Definition}

\author{
    \IEEEauthorblockN{V. Kulathumani\IEEEauthorrefmark{1}, A. Arora\IEEEauthorrefmark{2}, M. Sridharan\IEEEauthorrefmark{2}, K. Parker\IEEEauthorrefmark{2}, B. Lemon\IEEEauthorrefmark{1}} \\
    \IEEEauthorblockA{\IEEEauthorrefmark{1}West Virginia University
    \\\{vinod.kulathumani, bryan.lemon\}@mail.wvu.edu} \\
    \IEEEauthorblockA{\IEEEauthorrefmark{2}The Samraksh Company 
    \\\{anish.arora, mukunda.sridharan, kenneth.parker\}@samraksh.com}
}

\begin{document}
\title{On the repair time scaling wall for MANETs}

\maketitle

\begin{abstract}
The inability of practical MANET deployments to scale beyond about 100 nodes has traditionally been blamed on insufficient network capacity for supporting routing related control traffic. However, this paper points out that network capacity is significantly under-utilized by standard MANET routing algorithms at observed scaling limits. Therefore, as opposed to identifying the scaling limit for MANET routing from a capacity stand-point, it is instead characterized as a function of the interaction between dynamics of path failure (caused due to mobility) and path repair. This leads to the discovery of the repair time scaling wall, which is used to explain observed scaling limits in MANETs. The factors behind the repair time scaling wall are identified and techniques to extend the scaling limits are described.
\end{abstract}

\let\thefootnote\relax\footnotetext{The email point of contact is vinod.kulathumani@mail.wvu.edu. This work was supported in part by Defense Advanced Research Projects Agency (DARPA) contract FA8750-12-C-0278. The views, opinions, and/or findings contained in this paper are those of the authors and should not be interpreted as representing the official views or policies of the Department of Defense or the U.S. Government.}


\IEEEpeerreviewmaketitle

\section{Introduction}
Despite several years of research in MANETs, it turns out that most practical deployments do not scale beyond about 100 nodes \cite{rtsw1}. The most common reason attributed to this limitation is that of bounded wireless network capacity. As the number of nodes in a network grows, the overall capacity in the network only grows as $O(\sqrt{n})$ \cite{rtsw2}. Several researchers have compared this growth rate with the required capacity to support routing related control traffic (i.e., the network layer overhead), and have argued that the wireless network capacity does not scale with the routing needs. But our studies show that at a scale of about $75$-$150$ nodes, the amount of channel capacity used to support routing turns out to be under $2\%$, and yet the median path reachability (the number of connected paths) is less than $40\%$. 

Specifically, we constructed and simulated the standard implementation of the Optimized Link State Routing (OLSR) \cite{rtsw3} using network scenarios of size $75$-$150$ nodes. To calculate the number of broken paths, we took periodic snapshots of the nodes routing tables every $50$ milliseconds and traversed the path derived from the routing table to find the reachability of the nodes between every pair of nodes.  Further, in order to isolate the effect of Network Layer Overhead (NLO) on scaling, the simulations were done in the absence of any useful data traffic. At a scale of about $100$ nodes, we observed that around $60\%$ of the paths remained broken all the time. However, the network layer overhead was only accounting for about $2\%$ of the network capacity. The details of this simulation are described in Appendix A, but in summary, {\em our analysis shows that the routing system failed well before the capacity limits were reached}.

The objective of this paper is to investigate alternate reasons for this failure. To do so, instead of formulating this problem from a capacity standpoint, we characterize routing path reachability as a function of the interaction between the dynamics of path failure and path repair. Informally speaking (formal definitions are provided in Section IV), the path connectivity interval refers to the average time that end to end paths in a network remain connected before mobility causes the paths to be disconnected. Note that breaking of a single link on a path is sufficient for the path to break. On the other hand, the path repair interval refers to the distribution of repair times for end to end paths in a network. By comparing the path connectivity and repair intervals as a function of network size, we are able to identify factors other than capacity that limit the scaling of MANETs. Our analysis leads to several interesting findings which are summarized below.

\begin{itemize}
\item The median path connectivity interval falls as $O(1/\sqrt{n})$. where $n$ is the network size. On the other hand, the median path repair interval remains fairly constant irrespective of the network size and is roughly equal to the link failure estimation time. This is the time required to detect that a link no longer can be used for routing because a node has moved beyond its communication range. When the path connectivity interval falls below the path repair interval, we can expect a majority of the paths to be disconnected, and the scale at which this occurs is defined as the {\em repair time scaling wall} for the MANET.

\item The crucial controllable factor that impacts the repair time scaling wall is the link failure estimation delay. Using our analysis, one can determine the bound on link estimation delay for a given network size under a given mobility model so that the system stays within the repair time scaling wall. 

\item By analyzing the distribution of path repair intervals, we determine that most of the broken paths can be fixed quite close to the failed links and thus local updates are mostly sufficient for fixing broken paths and restoring connectivity. Therefore, contrary to previous knowledge, the channel capacity required for propagating link state updates throughout the network does not play the dominant role in limiting the scalability of MANETs. Instead, it is the convergence of path connectivity interval towards the path repair interval as network size increases, which plays a dominant role in limiting the scalability of MANETs. 

\end{itemize}

\noindent{\bf Outline of the paper:}~~In Section II, we state related work. In Section III, we state the network model. In Section IV and Section V, we separately analyze the path connectivity and path repair dynamics in a fully connected mobile ad-hoc network. We utilize the results of these analyses in Section VI, to characterize the repair time scaling wall for MANETs. We conclude in Section VII.

\section{Related Work}
\label{sec:related}
Our interest in analyzing point to point routing protocols for MANET stems from the necessity of these protocols for information sharing in MANETs as the scale of the networks starts to increase. Previous studies \cite{rtsw18} have compared the efficiency and impact of stateful communication strategies (such as point to point routing) and stateless strategies (such as flooding) for MANETs under different levels of network connectivity, density and mobility rates. The study in \cite{rtsw18} points out that under high mobility and connectivity, flooding is the right choice. But the study in \cite{rtsw18} does not consider network scale. While flooding based solutions may be acceptable for small scale networks, as the scale starts to increase to several hundreds and thousands of nodes, stateless protocols start becoming less of an alternative and point to point routing becomes necessary. Our point in this paper is that the scalability issues attributed to point to point routing are not caused by capacity constraints, but rather discovery latency. 

There has been plenty of research on routing algorithms for mobile, ad-hoc networks in the past two decades. Some well-known examples are OLSR \cite{rtsw3}, TBRPF \cite{rtsw4}, STAR \cite{rtsw5}, ZRP \cite{rtsw9}, DSR \cite{rtsw6}, AODV \cite{rtsw7}, and DSDV \cite{rtsw8}. At a high level, these algorithms can be classified in two ways: (i) proactive or reactive depending on whether they maintain up-to-date routing information at each node (proactive) or whether they determine routes on-demand when a packet is to be routed (reactive), and (ii) link-state or distance vector \cite{rtsw11}, depending on whether information about individual links is exchanged globally (link-state) or whether path information at each node is exchanged locally (distance-vector). 

In general, the main emphasis in these studies has been on reducing the network level overhead required for propagating route information \cite{rtsw19} so that they remain within the capacity scaling limits imposed in the seminal paper by Gupta and Kumar for wireless networks \cite{rtsw2}. While such a strategy may yield satisfactory routing performance under low mobility and in small networks, practical deployments have not been successful beyond 75-100 nodes \cite{rtsw1}. A case in point is the Optimized Link State Routing (OLSR) which introduces selection of multi-point relays (an optimally chosen subset of two-hop neighbors) for controlling the propagation of neighborhood information across the network. As a result, although the routing overhead is reduced, the quality of routing is poor, as quantified by our experimental results in the Appendix A of this paper. Moreover, at the point of poor routing performance, the capacity of the network is in fact severely underutilized. Therefore, our stand in this paper is that network layer overhead is not the right metric for characterizing routing performance and understanding scaling limits for MANETs. Instead, we formulate an alternate technique to quantify MANET scaling limits, wherein as opposed to analyzing routing protocols from a capacity standpoint, we analyze them by comparing the probability distributions of path connectivity and path repair intervals.

Our results in this paper show that it is not capacity but rather the latency in discovering broken links and fixing broken paths that imposes the crucial scaling limit for MANETs. We have shown this by characterizing, both analytically and experimentally, the dynamics of path connectivity and path repair in MANETs of different sizes. Our results point out that quick and efficient link estimation is critical for extending the scale of MANETs, which previous studies have largely overlooked. Another striking observation that we make in this paper is that the latency for fixing a majority of paths in the network is very close to the latency for discovering a broken link, implying that a majority of the paths can actually be fixed quite locally. This is significant because it shows that much of the routing overhead involved in propagating network wide information by traditional routing protocols is actually not needed. 
 
Lastly, we would like to point out that in this paper we are interested in real-time data routing and hence the idea of introducing delay tolerance to increase the network capacity and recent results on capacity-delay tradeoffs for mobile ad-hoc networks [20, 21] are not directly applicable here. Moreover, as soon as we weaken the notion of path connectivity by stretching temporally, our results do not apply.

\section{Model}
\label{sec:model}
We consider a mobile network of N nodes deployed over a two dimensional region. The communication range of the nodes is constant irrespective of network size N. For our analysis, we assume a random walk mobility model \cite{rtsw12} for the nodes (we later relax this assumption and evaluate our findings under multiple mobility models in ns-3 simulations). In the random walk mobility model, at each interval a node picks a random direction uniformly in the range $[0,2\pi]$ and moves with a constant speed randomly chosen in the range $[v_{min},v_{max}]$ for a constant distance $\phi$. At the end of each interval, a new direction and speed are calculated. This model is Brownian in its characteristics; the Brownian model can be described as a scaling limit of this motion model under small step sizes \cite{rtsw13}. The random walk motion model results in node locations that are uniformly distributed across the network \cite{rtsw14}. Therefore, we assume that over time the average number of neighbors per node is $\rho$ and this number stays constant irrespective of the network size.

We assume that the network is never partitioned. Thus, there exists at least one path between every pair of nodes in the network, at all times. The MANET is assumed to be supported by an underlying routing algorithm that is responsible for discovering a path between each pair of nodes, which is optimal under a given metric. Let $P_{(s,d)} (t) = <s,g_1,g_2,…,,g_k ,d>$  be the path between the pair of nodes $<s, d>$ as determined by the routing algorithm at time $t$. In this path, $g_1$ represents the next hop towards $d$ as determined by the routing table at node $s$, and $g_{i+1}$ represents the next hop towards as $d$ as determined by the routing table at node $g_i$. Since the network is mobile, neighboring nodes move in and out of each others' transmission range, thus adding and breaking links respectively. At a given network density and average node speed, the rate at which links are added and deleted are largely unaffected by network size and we denote these to be constants $z$ and $\theta$ respectively. The path $\gamma(s,d)$  at time $t$ is valid only if all the links along the paths exist at time $t$. Thus if any of the links are broken, the path is assumed to be disconnected. The underlying routing algorithm fixes such broken paths between node pairs, by first detecting broken paths and then restoring them via alternate routes. Thus, each end to end path in a MANET can be represented as a sequence of connected and disconnected intervals in time. 

\subsection{Routing and link estimation}

We assume that there is a correction process, i.e., a routing algorithm executing in the system. Specifically, we used an event-based link state routing algorithm (LSR) that generates a link state update per link estimation event (i.e., either the discovery of a new neighbor or the discovery of the loss of a neighbor). A beacon based algorithm is used for link estimation, i.e. each node beacons a heartbeat message at a steady rate of once every $B$ seconds. The radio range is kept constant to enable knowledge of ground truth. $3$ missed heartbeats are used to signal the loss of a neighbor. Each time a new or broken link is discovered at a node, the link state of the node is flooded to the entire network.

Note that we have deliberately avoided the use of optimization in the routing algorithm such as those implemented by OLSR for reducing the link state discovery overhead, for pruning the route dissemination graph and for reducing the amount of updates. By avoiding these optimization, we are able to clearly characterize the impact of link estimation times and route update times on the route reachability. In fact, one of our findings is that the optimization introduced by protocols such as OLSR may actually hurt the route reachability because they place a bound on how fast broken links can be discovered in the network.  

\section{Path failure dynamics}
\label{sec:pfd}
\subsection{Analytical characterization of path connectivity interval}

\begin{Definition}[Connectivity Interval of a given path]
Let $P_{x,y}(t)$ denote  the path between a pair of nodes $x$ and $y$ in the network at  a given time $t$. The connectivity interval for the path $P_{x,y}(t)$ is defined as the  duration after which at least one of the links on that path breaks, causing the path to break. 
\end{Definition}

Using the above definition, one can derive the average and median path connectivity intervals, computed over all paths in the network. 

\begin{Definition} [Average Path Failure Rate]
The average path failure rate is defined as the number of times that each end-to end path changes from a connected to disconnected state per unit time, averaged over the number of end-to-end paths in the network. Roughly speaking, the average path failure rate grows as the inverse of the average path connectivity interval.
\end{Definition}

\begin{theorem}
The average length of each path in a network of $n$ nodes is $O(\sqrt{n})$.
\end{theorem}

\begin{IEEEproof}
Consider any node in the network. The number of nodes within a circle of diameter d around the node is $O(\rho d^2)$, where $1 \leq d \leq \sqrt{n}$. Therefore, the number of nodes between a distance of $d$  and $(d-1)$ from point $p$ is $O(d)$, and the number of paths between lengths  $d$  and $(d-1)$ in the network is $O(nd)$. The total length of all $n^2$  paths in the network can be found as follows:

\begin{equation}
O(\sum\limits_{d=0}^{\sqrt{n}} (nd^2)) = O(n \sqrt{n}^3 )= O(n^{(5/2)}) \nonumber
\end{equation}

The average length of each path is thus equal to $O(\sqrt{n})$.
\end{IEEEproof}

\begin{theorem}
The average path connectivity interval for a MANET decays as  $O(1/\sqrt{n})$, where $n$ is the number of nodes in the MANET.
\end{theorem}

\begin{IEEEproof}
The link failure rate is a constant of $\theta$ per unit time. Thus the average interval between two successive times that a given link breaks is $\frac{1}{\theta}$. Now consider a set of $p$ links whose failures are uniformly distributed in time with a constant rate of $\theta$ per unit time. In this case, the average time between two successive link failures is $\frac{1}{\theta p}$ We note from the previous theorem that the average length of a path grows is $O(\sqrt{n})$ and that it is sufficient for one link on a path to break for the path to be disconnected. Thus, the average time that a path remains connected is $O(1/\theta \sqrt{n})$.
\end{IEEEproof}  

\subsection{Experimental Characterization of Path Connectivity Interval}

\subsubsection{Network setup}

We now characterize the path connectivity intervals for different network sizes using simulations in ns-3. In this section, we have considered a random walk 2-d mobility model where each node moves inside a fixed rectangular area with a speed chosen uniformly within the range $2-4$ m/s and changes directions after moving in a randomly chosen direction for $30m$.  The mobility model and average node speed are such that the number of link changes per second per node is approximately $0.25$ irrespective of the network size. The deployment area relative to the communication range is such that the average number of neighbors for each node is approximately $8$. Occasionally, the network may be disconnected because of mobility. our findings are validated for some other mobility models in Section $V$. 

\begin{figure*}[htbp]
\vspace*{-3mm}
  \begin{center}
    \mbox{
      \subfigure[] {\scalebox{0.35}{\includegraphics[width=\textwidth]{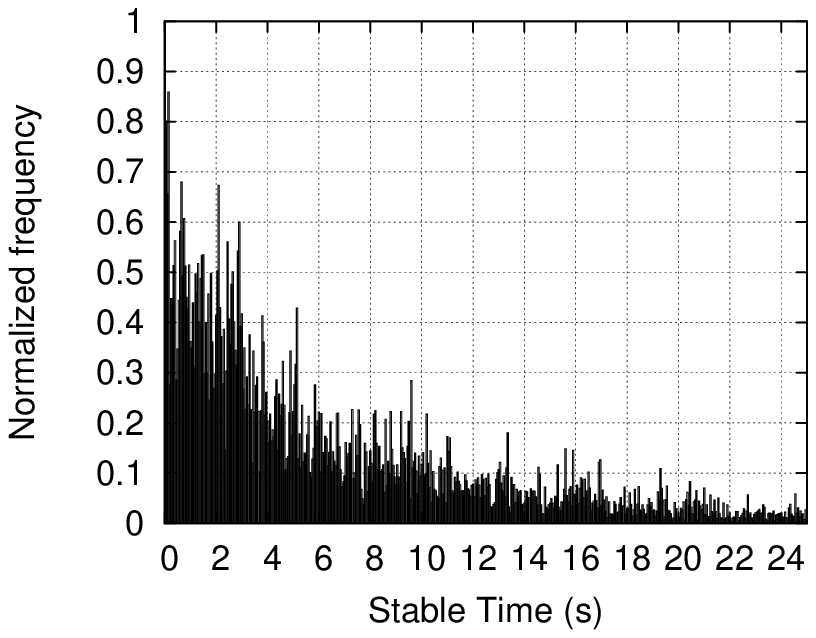}} \label{fig:1a}} \quad
      \subfigure[] {\scalebox{0.35}{\includegraphics[width=\textwidth]{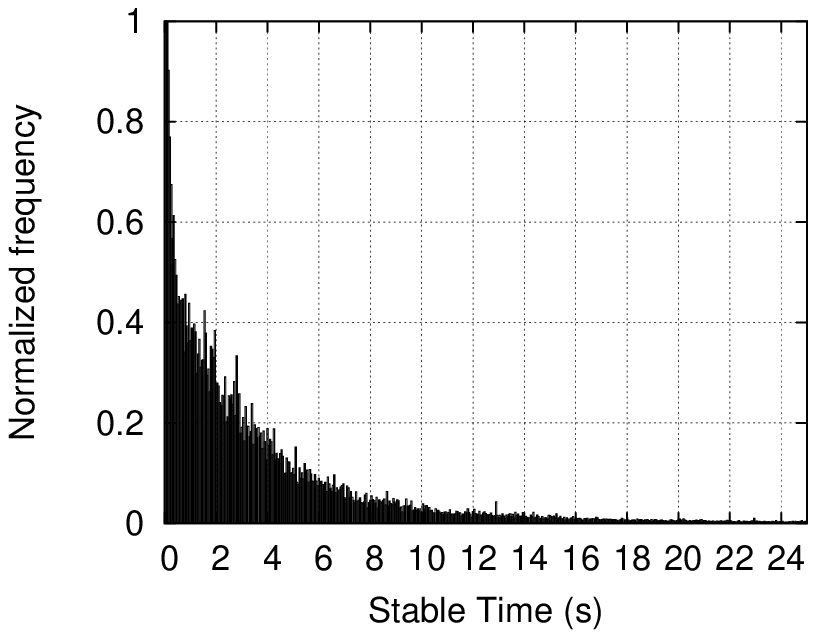}} \label{fig:1b}} \quad      
      } \\
    \mbox{
      \subfigure[] {\scalebox{0.35}{\includegraphics[width=\textwidth]{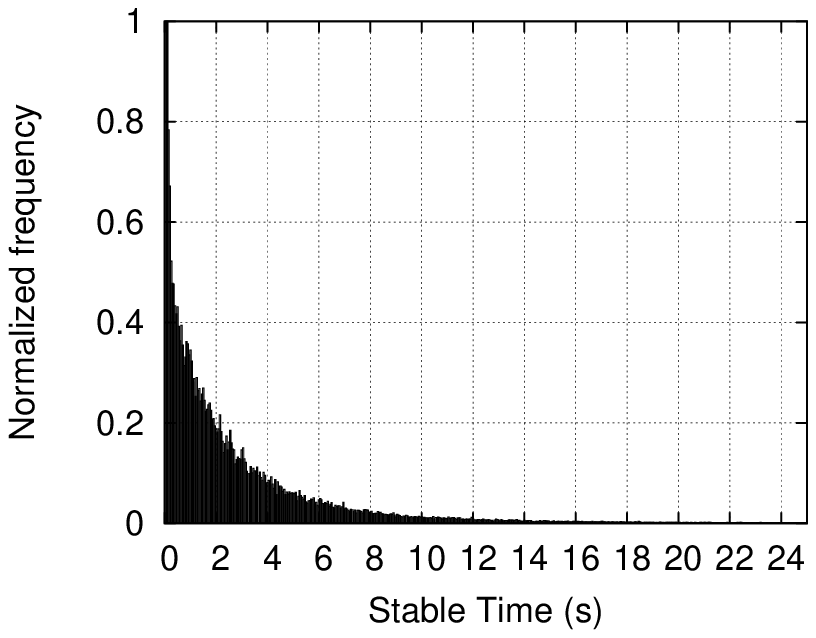}} \label{fig:1c}} \quad
      \subfigure[] {\scalebox{0.35}{\includegraphics[width=\textwidth]{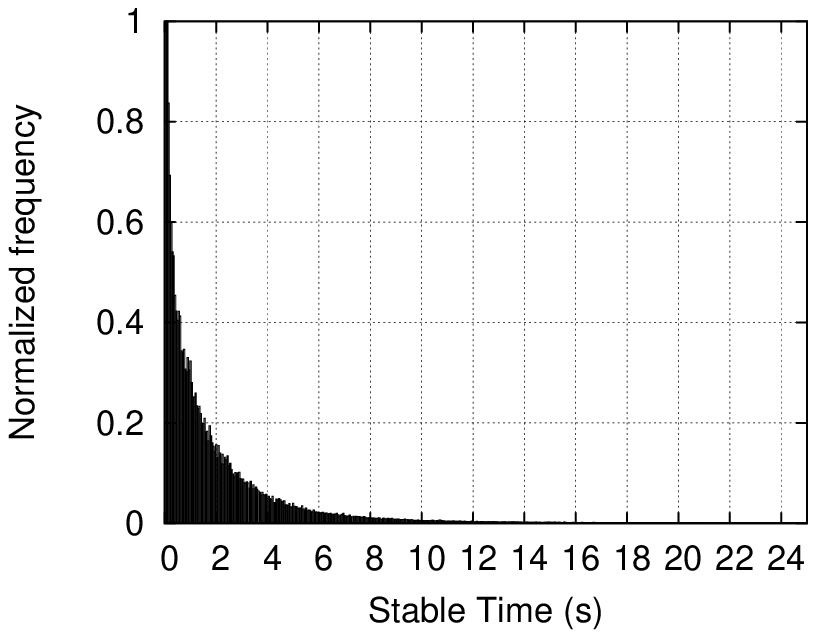}} \label{fig:1d}} \quad      
      }   
    \vspace{-2mm}  
    \caption{Histogram of Path Stability Intervals with 50, 150, 300 and 500 nodes. Note that the median and mean path stability intervals progressively shift to the left}
       \label{fig:rtsw1}
  \end{center}
\vspace*{-2mm}
\end{figure*}

\begin{figure*}[htbp]
\vspace*{-3mm}
  \begin{center}
    \mbox{
      \subfigure[] {\scalebox{0.4}{\includegraphics[width=\textwidth]{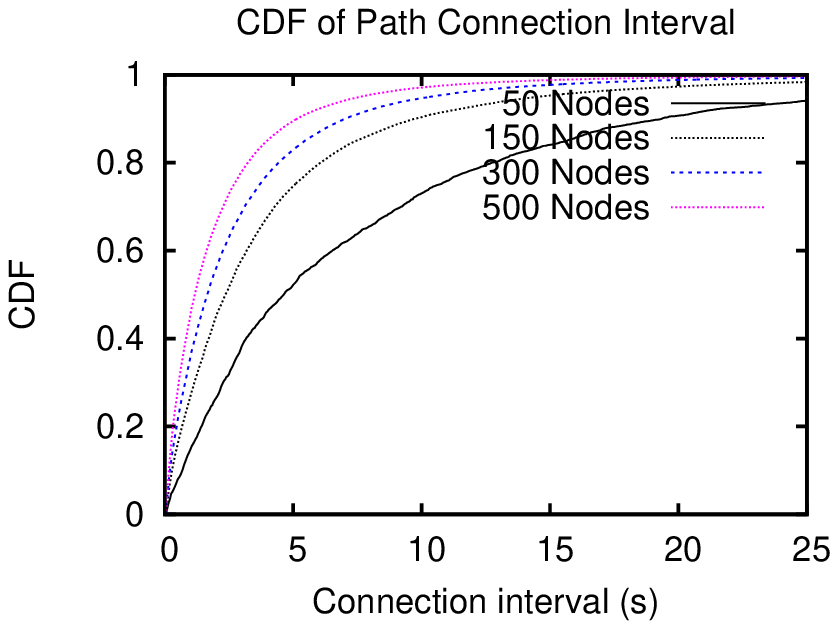}} \label{fig:2a}} \quad
      \subfigure[] {\scalebox{0.4}{\includegraphics[width=\textwidth]{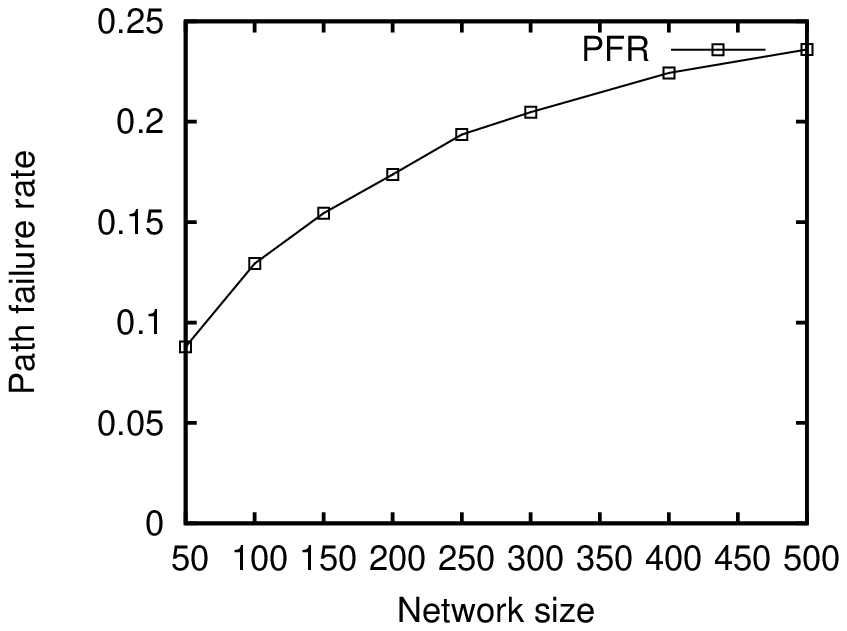}} \label{fig:2b}} \quad      
      } 
    \vspace{-2mm}  
    \caption{(a) Cumulative distribution function of path connectivity intervals, for different network sizes (b) Average path failure rate as a function of network size}
       \label{fig:rtsw2}
  \end{center}
\vspace*{-2mm}
\end{figure*}

\begin{figure*}[htbp]
\vspace*{-3mm}
  \begin{center}
    \mbox{
      \subfigure[] {\scalebox{0.4}{\includegraphics[width=\textwidth]{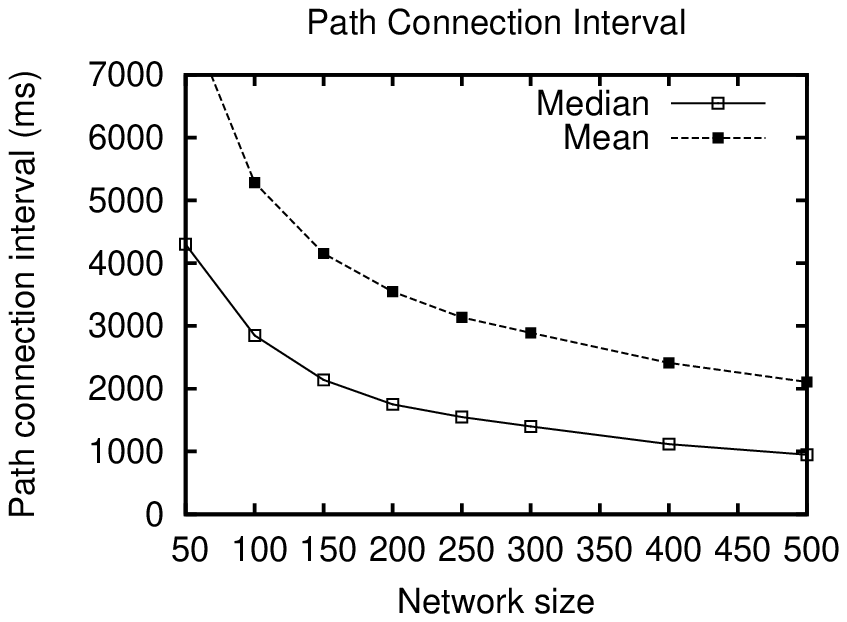}} \label{fig:3a}} \quad
      \subfigure[] {\scalebox{0.4}{\includegraphics[width=\textwidth]{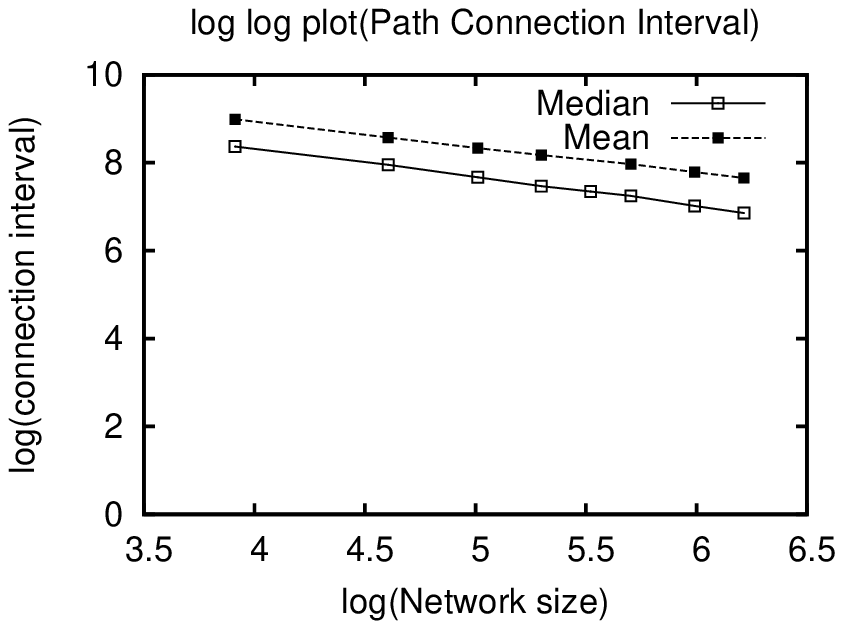}} \label{fig:3b}} \quad      
      } 
    \vspace{-2mm}  
    \caption{(a) Median and mean path connectivity interval as a function of network size. (b) log-log plot with a slope of approximately $-0.5$ highlight the $O(1/\sqrt{n})$ asymptotics}
       \label{fig:rtsw3}
  \end{center}
\vspace*{-2mm}
\end{figure*}

\subsubsection{Path connectivity results}
We collect routing statistics at regular intervals of $50$ ms during each trace of the simulation. This is done as follows. Between each pair of nodes $<s, d>$ in the network, $P_(s,d) (t) = <s,g_1,g_2,…,g_k,d>$  is the path as determined by the routing algorithm at time t. In this path, $g_{i+1}$ represents the next hop towards as $d$ as determined by the routing table at node $g_i$. We mark the path between $<s, d>$ as disconnected at time $t$, if any of the links along the path is not valid at time t. Link validity is determined by comparing against the ground truth location data and noting that a link between two nodes is invalid if the distance between them is greater than the radio range.

For each pair of nodes in the network, we note down the duration for which the path between them remains connected over the course of the simulation. Note that during the course of each simulation, paths will be continually disconnecting and getting repaired. Thus, the path connectivity interval for each instance of a path being connected is noted down. This data is used to plot the histogram of path connectivity intervals (shown in Fig.~\ref{fig:rtsw1}). The data in Fig.~\ref{fig:rtsw1} shows the histogram of path connectivity intervals at networks sizes ranging from $50$ to $500$ nodes. The y-axis is normalized by the total number of instances at each network size. We observe from this figure that the mean and median of the histograms shift progressively to the left. This indicates that the path connectivity intervals decrease with the network size. To better understand the histograms, in Fig.~\ref{fig:rtsw2} we plot the cumulative distribution function for the path connectivity intervals, computed over the duration of each simulation. This figure shows that the tail of the distribution becomes smaller as network size increases. 

Then we characterize the decay rate for the median and mean path connectivity intervals. Fig.~\ref{fig:rtsw3} shows that the median and mean path connectivity interval for the network decay as $O(1/\sqrt{n})$. Fig.~\ref{fig:3b} is a log-log plot of the path connectivity intervals. The slopes of the lines are approximately $-0.5$, highlighting the $O(1/\sqrt{n})$ asymptotics. The medians are lower than the respective means, indicating the large range for path connectivity intervals in the network. The connectivity intervals for some paths are relatively high.
 
In Fig.~\ref{fig:2b}, we show the average failure rate per path in the network. Specifically, we note the number of times that each path moves from the connected to disconnected state over the duration of the simulation, and average this over the number of paths in the network. Fig.~\ref{fig:2b} shows that the average path failure rate grows as $O(\sqrt{n})$. All these results match our analysis in Section IV.A. 

Finally, we note that during the simulations the network might be occasionally partitioned, i.e., there may be no paths in the ground truth between nodes at certain times. For computing the data shown in Fig.~\ref{fig:rtsw1}, Fig.~\ref{fig:rtsw2} and Fig.~\ref{fig:rtsw3}, we ignore such instances. Thus, the path connectivity intervals are only determined over pairs of nodes for which some path exists in the ground truth data

\section{Path repair dynamics}
\label{sec:prd}
Once a link state change has caused a path to disconnect, the repair process now consists of two components: (i) discovery of the failed link and (ii) propagation of the discovery to other nodes in the network so as to restore paths that were broken as a result of the link state change. In general, the updated link state information only needs to flood a small region completely surrounding the destination node in order to fix all broken routes. Once that has occurred packets originating from outside the region will follow the old routes until they intersect the region of updated routes, at which point they will follow the updated routes to the correct destination. This is what we call as the repair interval.

\begin{Definition}[Repair Interval of a given path]
The repair interval of a path between a pair of nodes $x$ and $y$, is the time taken to restore the connection between $x$ and $y$, starting from a disconnected state. 
\end{Definition}

Using the above definition, one can derive the average and median path repair intervals, computed over all paths in the network. {\em It is important to note that restoring connectivity does not mean that the path is now optimal and it does not mean that the path has stabilized.} As the information about link changes propagate in the network, better paths may continue to be established between the same pair of nodes. 

\subsection{Experimental Characterization of Path Repair Interval}

\begin{figure*}[htbp]
\vspace*{-3mm}
  \begin{center}
    \mbox{
      \subfigure[] {\scalebox{0.35}{\includegraphics[width=\textwidth]{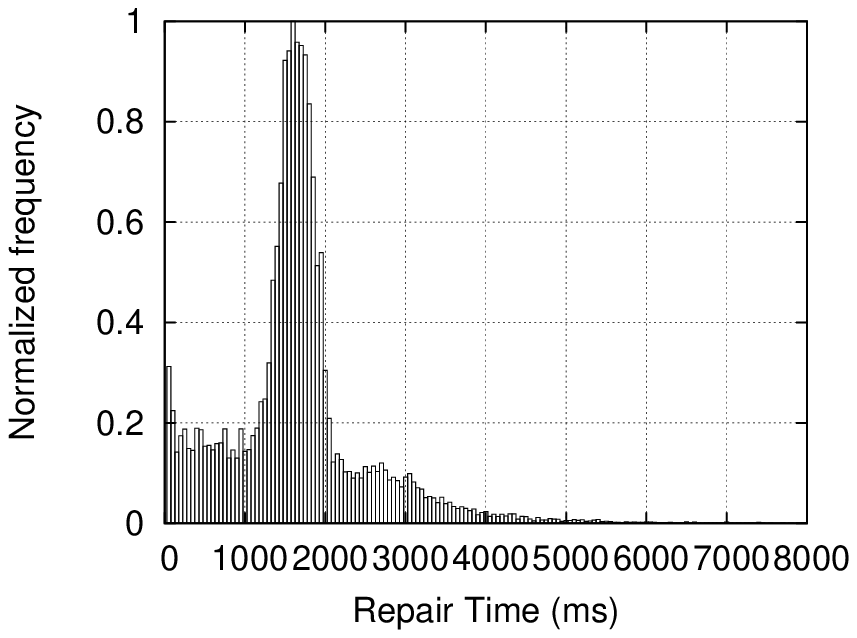}} \label{fig:4a}} \quad
      \subfigure[] {\scalebox{0.35}{\includegraphics[width=\textwidth]{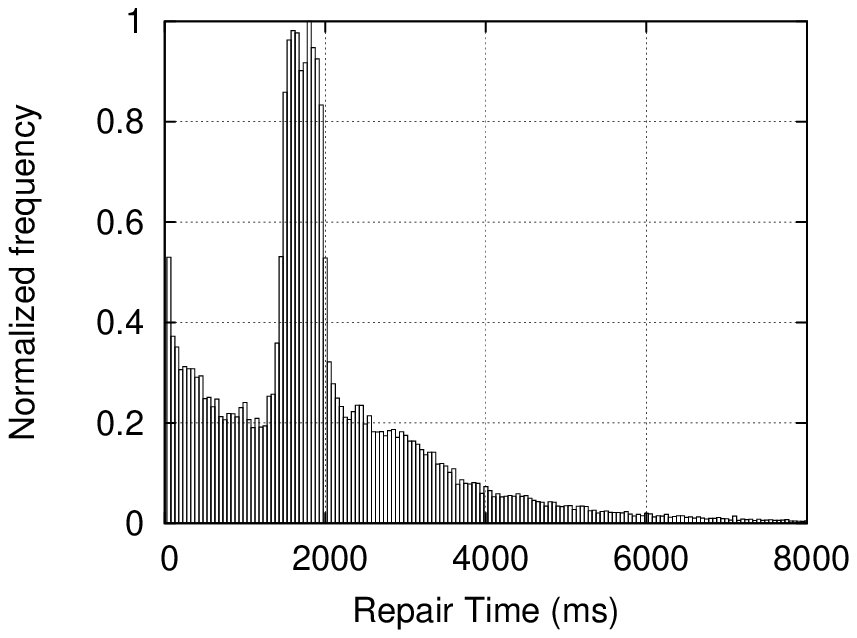}} \label{fig:4b}} \quad      
      } 
    \vspace{-2mm}  
    \caption{Histogram of Path Repair Intervals with 150 and 500 nodes. Note that the median path repair intervals stay approximately constant and equal to three times the heartbeat interval (i.e., the link failure estimation time)}
       \label{fig:rtsw4}
  \end{center}
\vspace*{-2mm}
\end{figure*}

\begin{figure*}[htbp]
\vspace*{-3mm}
  \begin{center}
    \mbox{
      \subfigure[] {\scalebox{0.4}{\includegraphics[width=\textwidth]{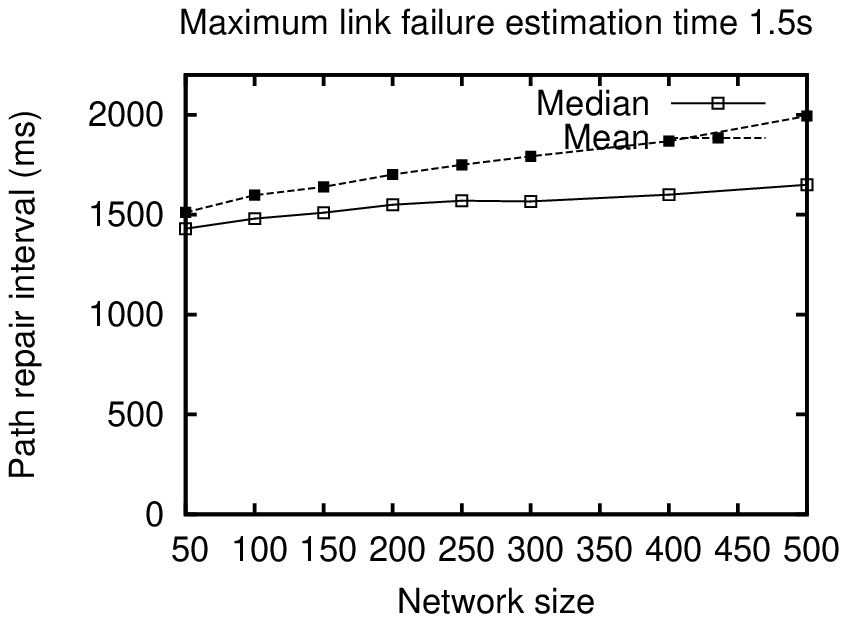}} \label{fig:5a}} \quad
      \subfigure[] {\scalebox{0.4}{\includegraphics[width=\textwidth]{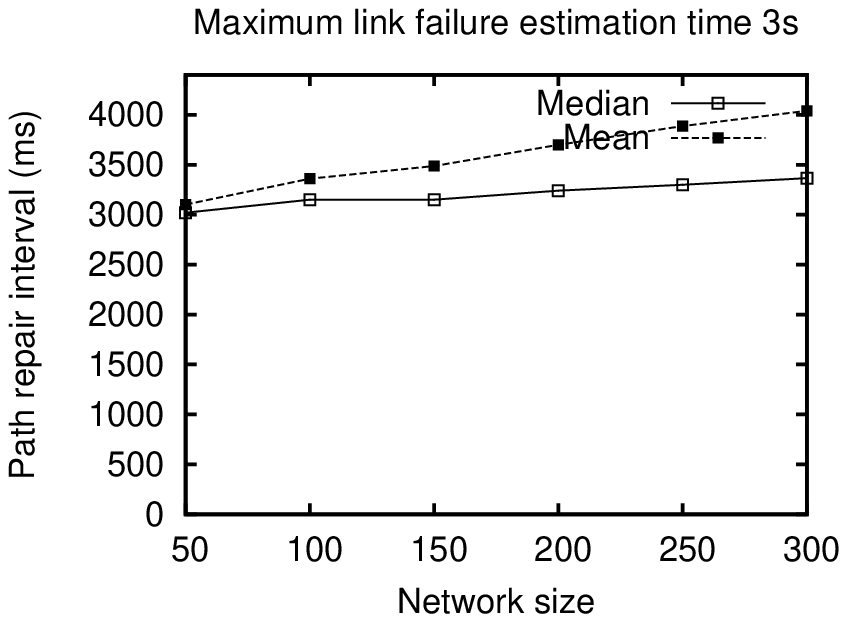}} \label{fig:5b}} \quad      
      } 
    \vspace{-2mm}  
    \caption{: (a) Median and mean path repair interval with a beacon interval of 500 ms (b) Median and mean path repair interval with a beacon interval of 1000 ms. The plots show that the median repair interval stays approximately equal to (slightly higher than) three times the heartbeat interval, i.e., roughly equal to the time that it takes to discover a failed link.}
       \label{fig:rtsw5}
  \end{center}
\vspace*{-2mm}
\end{figure*}

As described in Section IV.A, we collect routing statistics at regular intervals of 50ms during each trace of the simulation. For each pair of nodes in the network, we note down the duration for which the path between them remains disconnected (i.e., under repair) over the course of the simulation. Note that during the course of each simulation, paths will be continually disconnecting and getting repaired. The path repair interval for each instance of a path being connected is noted down. This data is used to plot the histogram of path repair intervals (shown in Fig.~\ref{fig:rtsw4}). The y-axis is normalized by the total number of instances at each network size. As stated in Section IV.A, we ignore node pairs for which no path exists in the ground truth data. 

We observe from Fig.~\ref{fig:rtsw4} that the median path repair intervals stay approximately constant. This is highlighted more clearly in Fig.~\ref{fig:rtsw5}, where the median and mean repair intervals are plotted as function of the network size. This data is shown at two different heartbeat intervals (500ms and 1000ms). Note that this corresponds to link failure estimation times of 1500ms and 3000ms respectively (i.e., it takes 3 missed heartbeats to signal that a link does not exist). We make 3 observations based on this figure.

\begin{itemize}
\item	First, we see that the repair intervals follow a uniform distribution until the points close to the median. Then we notice that most of the paths have repair interval close to the median and the repair intervals follow a power law distribution after this point and quickly taper off. 

\item	Second, we observe that the median and mean exhibit a small increase with the network size, but this increase is more pronounced for the mean. This is because as network size increases, the worst case repair intervals tend to increase. For some paths, the discovery of a failed link needs to be propagated throughout the network before the path can be fixed.

\item	The median repair interval stays approximately equal to (slightly higher than) three times the heartbeat interval, i.e. roughly equal to the time that it takes to discover a failed link. Thus for a majority of the paths, the repair interval is close to the link failure estimation time. This indicates that a majority of the paths are fixed close to the location where a link failure is discovered, and that {\em the most significant factor in path repair is the time that it takes to detect a failed link}.
\end{itemize}

\section{Joint analysis of path failure and path repair}
\label{sec:joint}
By putting together the analysis of path stability and path repair dynamics, we are able to make some interesting observations on the scaling limits for MANETs. We describe these results in this section.

\subsection{Existence of repair time scaling wall}

\begin{figure*}[htbp]
\vspace*{-3mm}
  \begin{center}
    \mbox{
      \subfigure[] {\scalebox{0.4}{\includegraphics[width=\textwidth]{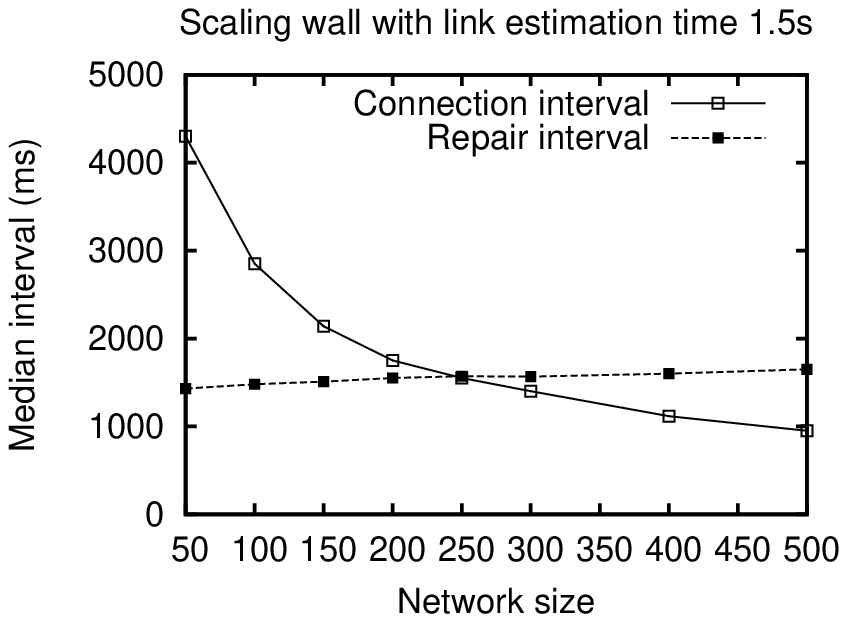}} \label{fig:6a}} \quad
      \subfigure[] {\scalebox{0.4}{\includegraphics[width=\textwidth]{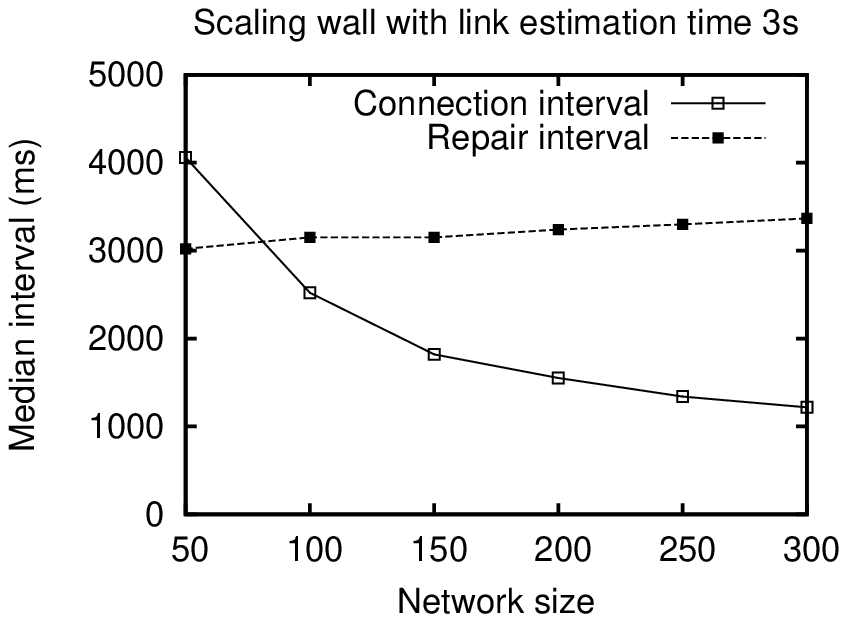}} \label{fig:6b}} \quad      
      } 
    \vspace{-2mm}  
    \caption{: (a) Repair time scaling is wall is around 300 nodes with a link estimation beacon interval of 500ms, i.e. link failure estimation time of 1500ms (b) Repair time scaling is wall is around 75 nodes with a link estimation beacon interval of 1000ms, i.e. link failure estimation time of 3000ms}
       \label{fig:rtsw6}
  \end{center}
\vspace*{-2mm}
\end{figure*}

In general, we expect good connectivity in the network when paths are repaired much faster than the rate at which they are broken. Based on this idea, we have defined repair time scaling wall in this paper as the point at which the median path connectivity interval falls below the median path repair interval. Other definitions for the repair time scaling wall are possible by considering different comparison points between the path connectivity interval and path repair interval. We consider this particular definition of the scaling wall to be significant because we expect that at this point a majority of data packets being routed through the network are expected to almost always encounter a broken path somewhere along their path. In systems where intermediate nodes drop data packets upon reaching a dead-end, this would imply poor throughput. In systems where intermediate nodes buffer dropped packets and re-establish a route, this would imply higher latency and a higher buffering overhead.

\begin{figure*}[htbp]
\vspace*{-3mm}
  \begin{center}
    \mbox{
      \subfigure[] {\scalebox{0.4}{\includegraphics[width=\textwidth]{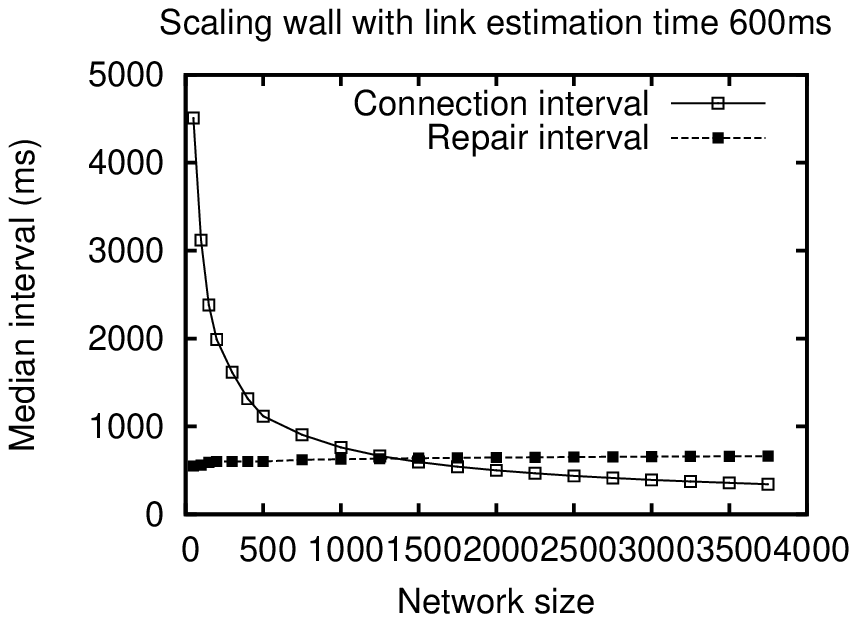}} \label{fig:7a}} \quad
      \subfigure[] {\scalebox{0.4}{\includegraphics[width=\textwidth]{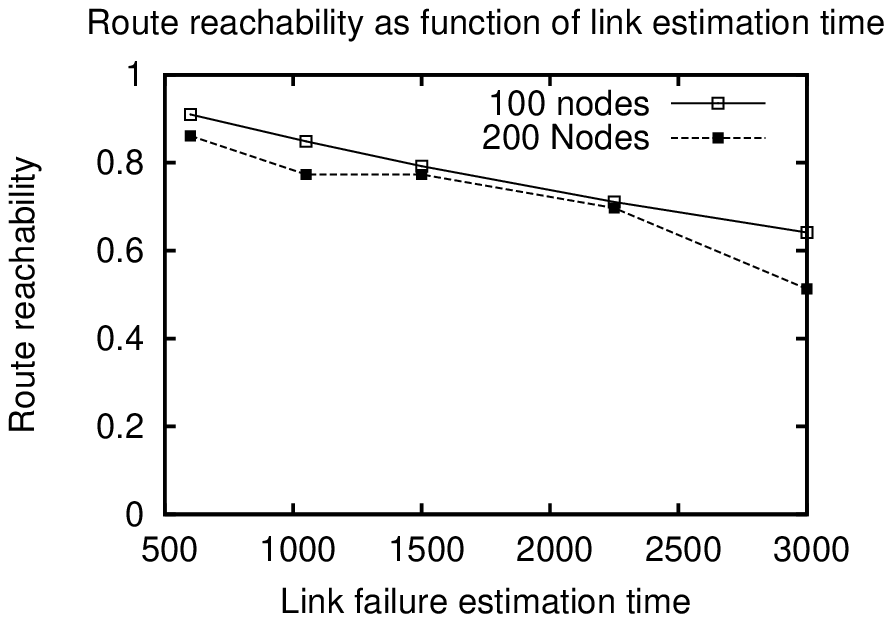}} \label{fig:7b}} \quad      
      } 
    \vspace{-2mm}  
    \caption{: (a) Repair time scaling wall is pushed to 1500 nodes with a link estimation beacon interval of 200ms (5Hz), i.e., link failure estimation time of 600ms (b) The percentage of reachable routes improves as link failure estimation time decreases}
       \label{fig:rtsw7}
  \end{center}
\vspace*{-2mm}
\end{figure*}

For a given mobility model and repair interval, our individual analysis of path connectivity and path repair dynamics can be used to determine the repair time scaling wall. This is shown in Fig.~\ref{fig:rtsw6}. In this figure, the path connectivity interval and path repair interval are compared with two different link estimation parameters, one with a heartbeat rate of 500ms and another with a heartbeat rate of 1000ms, for the random walk 2-d mobility model. The repair time scaling wall is indicated for both these scenarios as the intersection of the path repair and path connectivity interval. With a heartbeat frequency of 500ms, the expected scaling limit is about 300 nodes. With a heartbeat frequency of 1s (i.e., a link failure estimation time of approximately 3s), the expected scaling limit is only about 75-100 nodes. 

\subsection{Factors behind the scaling wall}

The crucial controllable factor that impacts the stabilization scaling wall is the link failure estimation time. The faster the failures can be detected the lower will be the expected repair interval – thus increasing the scalability of the system. This highlights the importance of faster link estimation in a MANET, especially that of faster link {\em failure} estimation. 
	
By decreasing the link failure estimation time, the scaling wall can be pushed farther. In Fig.~\ref{fig:7a}, we show the scaling wall with a heartbeat frequency of 200ms (5Hz). The link failure estimation time (and hence the median repair interval) is approximately around 600ms. This pushes the scaling wall to approximately 1500 nodes. Note that the path connectivity intervals only depend on the network size, and are independent of the heartbeat frequency or the link estimation time. 

In Fig.~\ref{fig:7b}, we analyze the impact of link estimation on route reachability. We count the percentage of routes that are connected in the network at intervals of 50ms and then average that over the duration of the simulation. Fig.~\ref{fig:6b} shows that route reachability steadily improves with faster link estimation. 

Our analysis of the path connectivity intervals can also be used to determine bounds on the path repair interval, such that the system does not cross the repair time scaling wall for any given network size. For instance, we can observe from Fig.~\ref{fig:7a} that for a network size of 2500 nodes, the path repair interval should be lower than 400ms. Such a determination can be used to guide the design of the link estimation strategy and parameters such as beaconing interval, duty cycling etc.

While the link estimation delay can be decreased by faster beaconing, we note that the scheduling of these beacons will impose a lower bound on the achievable link estimation. In the duty cycled, almost always-off scenario, the lower bound will be significantly higher.

\subsection{Verification on OLSR}

\begin{figure*}[htbp]
\vspace*{-3mm}
  \begin{center}
    \includegraphics[width=\textwidth]{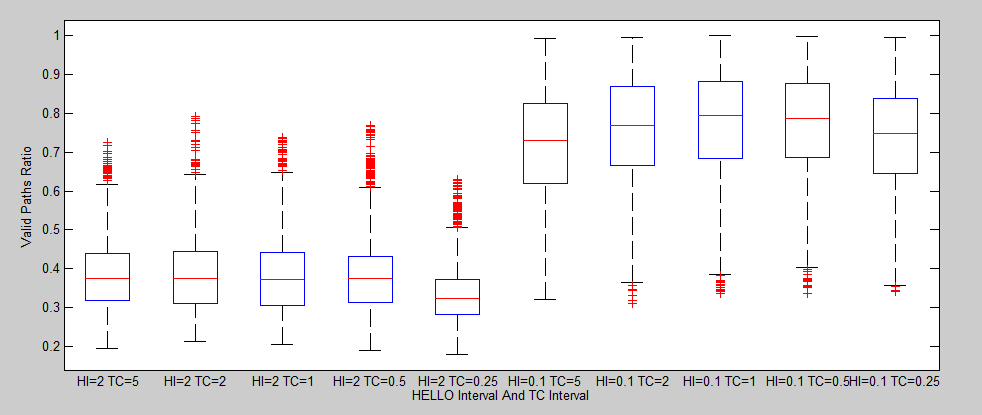} 
    \vspace{-2mm}  
    \caption{Impact of TC interval on route reacability in OLSR at two different Hello intervals (0.1s and 2s). TC interval does not have significant impact on route reachability. }
       \label{fig:rtsw8}
  \end{center}
\vspace*{-2mm}
\end{figure*}

Thus far, we have analyzed the route repair dynamics using a simple event based routing algorithm as described in Section~III. We have deliberately avoided the use of optimization in the routing algorithm such as those implemented by OLSR for reducing the link state discovery overhead, and network layer flooding overhead. By avoiding these optimization, we have been able to clearly characterize the impact of link estimation times and route update times on the route reachability. We now verify that our findings on rote reachability are valid even for OLSR.

\begin{figure}[htbp]
\vspace*{-3mm}
  \begin{center}
    \includegraphics[width=0.45\textwidth]{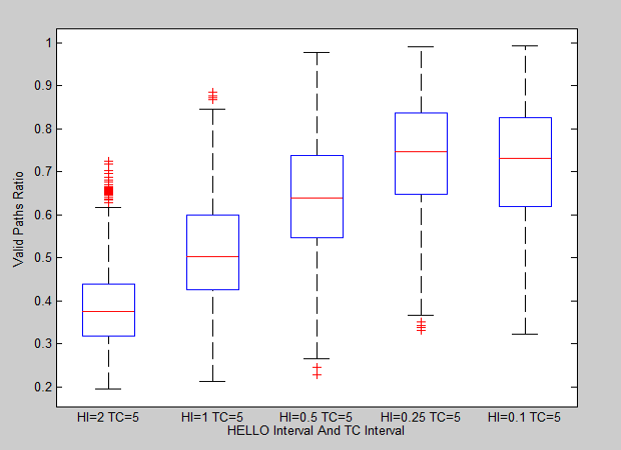} 
    \vspace{-2mm}  
    \caption{Impact of Hello on route reacability in OLSR. }
       \label{fig:rtsw9}
  \end{center}
\vspace*{-2mm}
\end{figure}

OLSR uses two kinds of messages: (i) Hello messages are used as neighborhood discovery beacons and are single hop broadcast messages (ii)TC or topology control messages are used to optimally flood the link state information in the network. A two layer hierarchy is used in the form of multi-point relays (MPRs) for optimally flooding the link state information. In Fig.~\ref{fig:rtsw9}, we show the route reachability in OLSR as a function of the Hello interval in a network of $100$ nodes (computed exactly as it was for the plain link state routing algorithm). We notice a similar improvement in route reachability even in OLSR, although the maximum is somewhat lower than plain LSR (i.e., without hierarchy and optimizations). Fig.~\ref{fig:rtsw8} emphasizes our findings even more, where we show the route reachability as a function of TC interval for two different Hello intervals. {\em We observe that increasing the rate at which information is flooded through the network does not have a significant impact, while faster link estimation does.}

\subsection{Verification with alternate mobility models}

\begin{figure*}[htbp]
\vspace*{-3mm}
  \begin{center}
    \mbox{
      \subfigure[] {\scalebox{0.32}{\includegraphics[width=\textwidth]{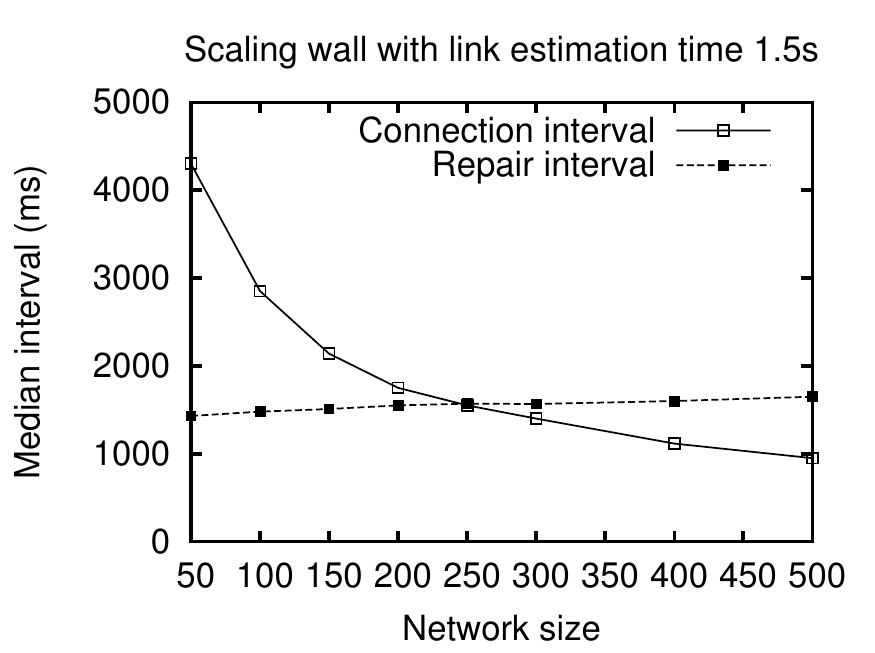}} \label{fig:10a}} \quad
      \subfigure[] {\scalebox{0.32}{\includegraphics[width=\textwidth]{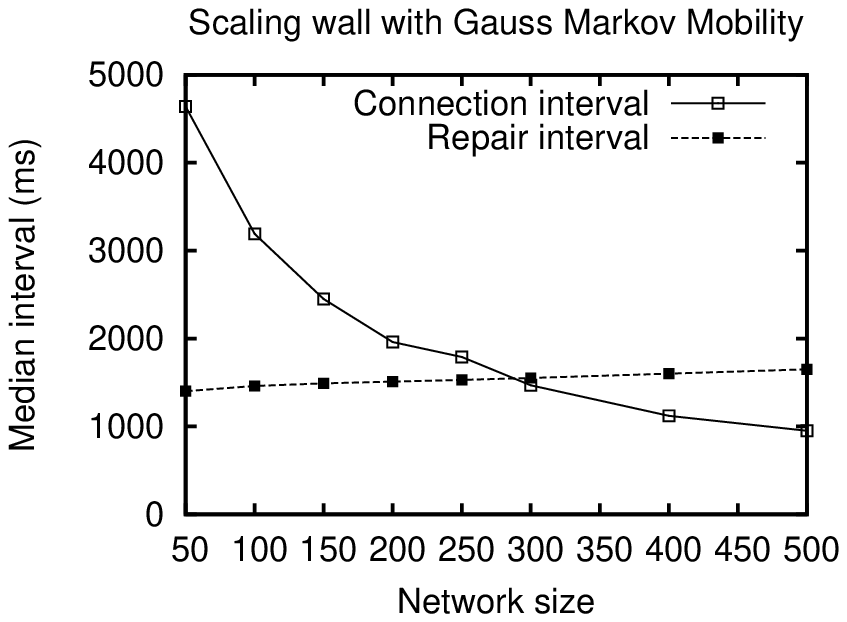}} \label{fig:10b}} \quad
      \subfigure[] {\scalebox{0.32}{\includegraphics[width=\textwidth]{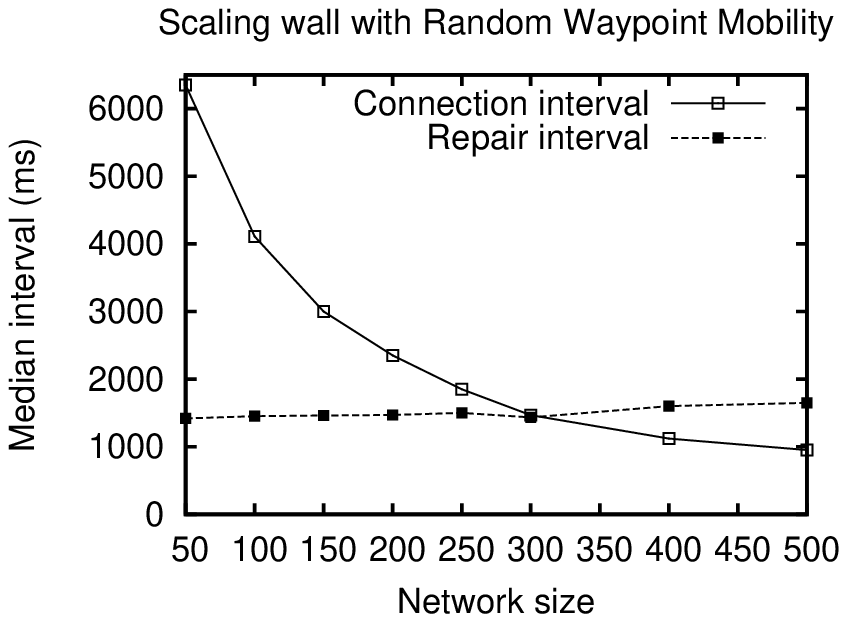}} \label{fig:10c}} \quad           
      } 
    \vspace{-2mm}  
    \caption{ Repair time scaling wall with a link failure estimation time of 1500ms under (a) random walk 2-d mobility model (b) Gauss-Markov mobility model and (c) random waypoint mobility model}
       \label{fig:rtsw10}
  \end{center}
\vspace*{-2mm}
\end{figure*}

While our analysis thus far was using the random walk 2-d mobility model, in this section we show the repair time scaling wall on two other mobility models, namely random waypoint and Gauss Markov. The node speeds remain in the range of $2-4$ m/s. For random way point, the pause time is set to $2$ seconds between successive changes. In the Gauss Markov model, where motion characteristics are correlated with time, tuned with a parameter $\alpha$. we have set $\alpha=0.75$. Velocity and direction are changed every $1$ second in the Gauss Markov Model. In Fig.~\ref{fig:rtsw10}, we compare the repair time scaling wall across the different mobility models with a link failure estimation time of 1500ms and observe that the scaling wall remains close to $300$ nodes for all these models.

\subsection{Impact of node speed and link dynamics}

\begin{figure*}[htbp]
\vspace*{-3mm}
  \begin{center}
    \mbox{
      \subfigure[] {\scalebox{0.32}{\includegraphics[width=\textwidth]{eps/joint500ms}} \label{fig:11a}} \quad
      \subfigure[] {\scalebox{0.32}{\includegraphics[width=\textwidth]{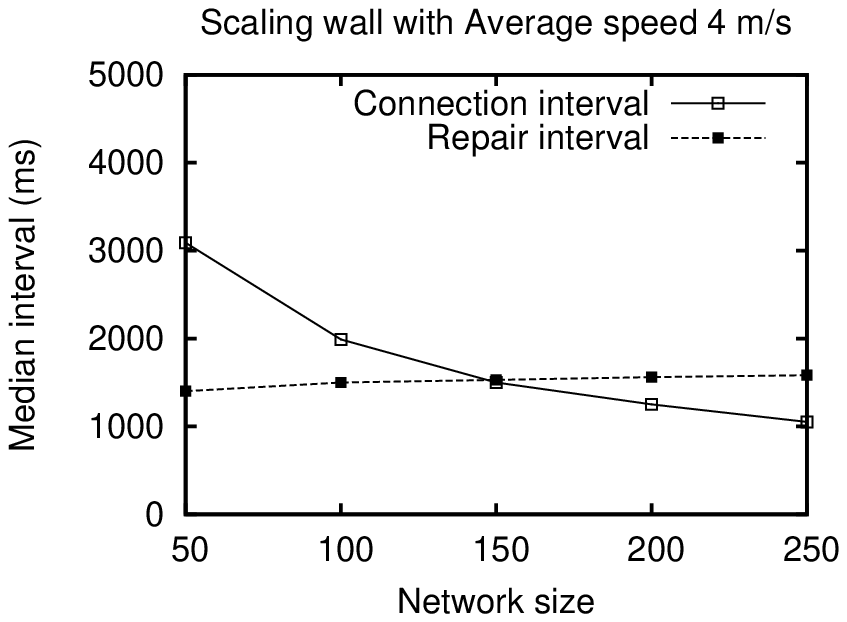}} \label{fig:11b}} \quad
      \subfigure[] {\scalebox{0.32}{\includegraphics[width=\textwidth]{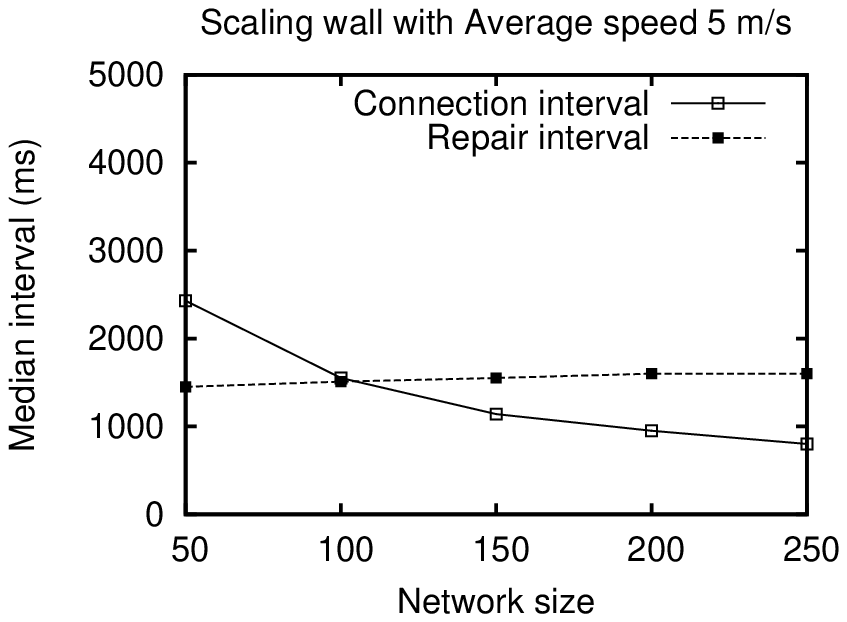}} \label{fig:11c}} \quad           
      } 
    \vspace{-2mm}  
    \caption{ Repair time scaling wall with a link failure estimation time of $1500$ms under (a) speeds in the range $2$ to $4$ m/s (b) speeds in the range $3$ to $5$ m/s and (c) speeds in the range $4$ to $6$ m/s}
       \label{fig:rtsw11}
  \end{center}
\vspace*{-2mm}
\end{figure*}

\begin{figure}[htbp]
\vspace*{-3mm}
  \begin{center}
    \includegraphics[width=0.45\textwidth]{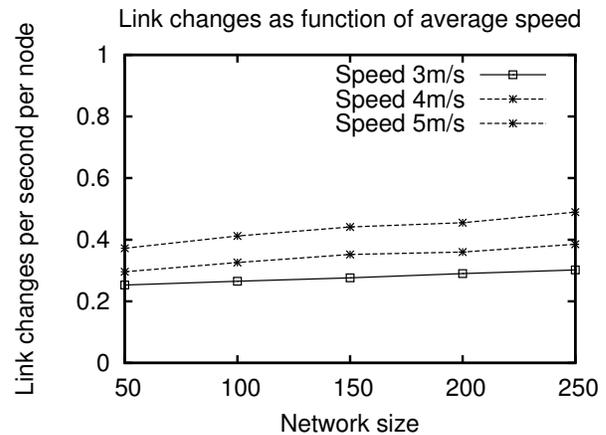} 
    \vspace{-2mm}  
    \caption{Link dynamics as a function of network size and node speeds }
       \label{fig:rtsw12}
  \end{center}
\vspace*{-2mm}
\end{figure}

We now analyze the impact of node speed on the repair time scaling wall. When node speed increases, link dynamics become faster and as a result the median path connectivity interval decreases. this results in a smaller repair time scaling wall as indicated in Fig.~\ref{fig:rtsw11}. The average node speeds are $3$, $5$ and $7$ m/s respectively. In Fig.~\ref{fig:rtsw12}, we have characterized the impact of these node speeds on link dynamics by showing the average number of link changes per second per node in the network. The figure shows that link change rate has almost doubled as the average speed increases from $3$m/s to $5$m/s. However the link dynamics remain relatively unchanged with network size. 

\section{Conclusions}
\label{sec:discussion}
In this paper, we have identified an important factor that limits scalability of routing systems in MANETs, namely the repair time scaling wall. The repair time scaling wall occurs because as network size increases, the average duration that a path remains connected decreases, while the average duration to repair a path remains fairly constant. When the average path connectivity interval falls below the path repair interval, the scaling wall is reached. We have shown that the path repair interval is roughly equal to the link failure estimation delay in the system. Thus, faster link estimation is critical for extending the path stability scaling wall - more specifically the time to estimate link failures. This leads us to explore efficient techniques for link estimation and to identify the limits on link estimation intervals, which is a subject of our ongoing work. 

Our analysis of the distribution of path repair intervals shows that a majority of the paths have repair time almost equal to the link failure estimation time and the repair intervals fall off as a power law distribution after the median. This indicates that very few paths require large, multi-hop information propagation before connectivity is restored. In general, local updates are mostly sufficient for fixing broken paths and restoring connectivity. This is actually good news from a scalability standpoint. This is because link estimation by itself only requires local information exchange and is not capacity intensive.


Our analysis also shows that much of the channel capacity utilized by standard versions of LSR in propagating link state updates throughout the network do not yield significant improvements in route reachability. This leads us to explore alternate routing protocols, where link state updates are only conditionally forwarded based on their expected impact on the path changes. This is also a subject of our ongoing work. While it is true that such a protocol may not always correct all the paths, need to know LSR can be supplemented by low frequency link state updates that propagate throughout the network.

In this paper, we have compared the path connectivity intervals with the path repair intervals. However, we note that even when a path stays connected, the paths may fluctuate. Some of these could be genuine fluctuations in search of better (optimal) paths while others may be a result of false link estimation events generated by the link estimation service. Such unnecessary fluctuations may have a cascading impact on the system performance because each link state event results in control traffic for route correction. Therefore, the stability of connected paths is an important metric that needs to be studied further.

\bibliographystyle{plain}
\bibliography{vinod}

\appendix
In this section, we describe our experiments on the standard implementation of the Optimized Link State Routing (OLSR) to study its documented poor performance \cite{rtsw1} in network sizes of 75-150 nodes. We used the default parameters of OLSR with a hello interval of 2 seconds, and topology control interval of 5 seconds.

We used the same network setup as described in Section~IV.B for our evaluations, i.e, a random walk 2-d mobility model with node speeds of $2$-$4$ m/s.  Further, in order to isolate the effect of Network Layer Overhead (NLO) on scaling, the simulations were done in the absence of any useful data traffic. The expected result of these simulations was that when the network size reached around ~100 nodes, the system would start to fail and the NLO at that point would have reached between 15\%-30\% of the network capacity, thereby leaving very little capacity for useful data traffic. Thus, our definition of the capacity wall for these simulations was NLO reaching 15\%-20\% of channel capacity.

\begin{figure}[htbp]
\vspace*{-3mm}
  \begin{center}
    \includegraphics[width=0.48\textwidth]{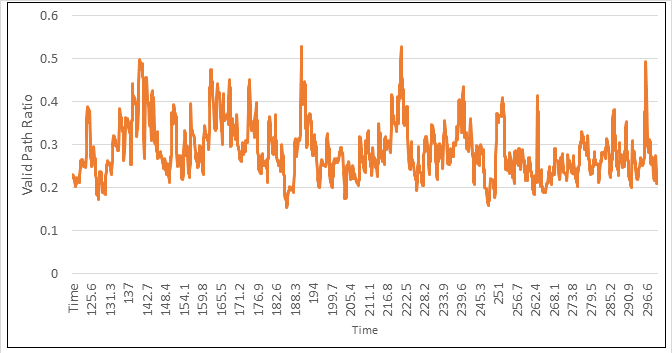} 
    \vspace{-2mm}  
    \caption{Poor route reachability ($<40\%$) in OLSR at a scale of 100 nodes; network layer overhead is less than $1$\% of the network capacity.}
       \label{fig:rtsw13}
  \end{center}
\vspace*{-2mm}
\end{figure}

Counter to our expectations, for a simulation of 100 nodes the OLSR network layer overhead did not consume around $15\%$ of the capacity. In fact, the NLO was accounting for around $1\%$ of the physical capacity. But further investigation revealed that the routing performance in the network was indeed bad. The real problem turned out to be the percentage of broken paths. OLSR in fact was minimizing NLO at the expense of routing performance. To calculate the number of broken paths, we took periodic snapshots of the nodes routing tables every 50 ms and traversed the path derived from the routing table to find the reachability of the nodes between every pair of nodes. Fig.~\ref{fig:rtsw13} shows the plot of the percentage of reachable paths for OLSR over a duration of $3$ minute simulation time. From the figure we observe that around $60\%$ of the paths remain broken almost throughout the entire period of the simulation. These results show that the routing system failed well before the capacity limits were reached.

\end{document}